\documentclass{article}

\usepackage[numbers]{natbib}         
\usepackage[colorlinks]{hyperref}    
\usepackage[english]{babel} 
\usepackage{amssymb}
\usepackage{amsmath}
\usepackage{txfonts}
\usepackage{mathdots}
\usepackage[classicReIm]{kpfonts}
\usepackage[dvips]{graphicx} 
\usepackage[a4paper, portrait, margin=1in]{geometry}
\usepackage{tabularx}
\usepackage{longtable}
\usepackage{multirow}
\usepackage{booktabs}
\usepackage[labelsep=period]{caption}
\usepackage{makecell}
\begin{document}
	\setlength{\parindent}{0pt}
	\setlength{\parskip}{1ex}
	
	\textbf{\Large Synthetic MRI-aided Head-and-Neck Organs-at-Risk Auto-Delineation for CBCT-guided Adaptive Radiotherapy}
	
	\bigbreak

	Xianjin Dai$^1$, Yang Lei$^1$, Tonghe Wang$^{1,2}$, Anees H. Dhabaan$^{1,2}$, Mark McDonald$^{1,2}$, Jonathan J. Beitler$^{1,2}$, Walter J. Curran$^{1,2}$, Jun Zhou$^{1,2}$, Tian Liu$^{1,2}$ and Xiaofeng Yang$^{1,2}$*
	
	$^1$Department of Radiation Oncology, Emory University, Atlanta, GA 30322
	
	$^2$Winship Cancer Institute, Emory University, Atlanta, GA 30322

	\bigbreak
	\bigbreak
	\bigbreak

	\textbf{*Corresponding author: }
	
	Xiaofeng Yang, PhD
	
	Department of Radiation Oncology
	
	Emory University School of Medicine
	
	1365 Clifton Road NE
	
	Atlanta, GA 30322
	
	E-mail: xiaofeng.yang@emory.edu

	\bigbreak
	\bigbreak
	\bigbreak
	\bigbreak
	\bigbreak
	\bigbreak

	\textbf{Abstract}

	\textbf{Purpose:} Organ-at-risk (OAR) delineation is a key step for cone-beam CT (CBCT) based adaptive radiotherapy planning that can be a time-consuming, labor-intensive, and subject-to-variability process. We aim to develop a fully automated approach aided by synthetic MRI for rapid and accurate CBCT multi-organ contouring in head-and-neck (HN) cancer patients.
	
	\textbf{Method:} MRI has superb soft-tissue contrasts, while CBCT offers bony-structure contrasts. Using the complementary information provided by MRI and CBCT is expected to enable accurate multi-organ segmentation in HN cancer patients. In our proposed method, MR images are firstly synthesized using a pre-trained cycle-consistent generative adversarial network given CBCT. The features of CBCT and synthetic MRI are then extracted using dual pyramid networks for final delineation of organs. CBCT images and their corresponding manual contours were used as pairs to train and test the proposed model. Quantitative metrics including Dice similarity coefficient (DSC), Hausdorff distance 95\% (HD95), mean surface distance (MSD), and residual mean square distance (RMS) were used to evaluate the proposed method.  
	
	\textbf{Result:} The proposed method was evaluated on a cohort of 65 HN cancer patients. CBCT images were collected from those patients who received proton therapy. Overall, DSC values of 0.87 ± 0.03, 0.79 ± 0.10 / 0.79 ± 0.11, 0.89 ± 0.08 / 0.89 ± 0.07, 0.90 ± 0.08, 0.75 ± 0.06 / 0.77 ± 0.06, 0.86 ± 0.13, 0.66 ± 0.14, 0.78 ± 0.05 / 0.77 ± 0.04, 0.96 ± 0.04, 0.89 ± 0.04 / 0.89 ± 0.04, 0.83 ± 0.02, and 0.84 ± 0.07 for commonly used OARs for treatment planning including brain stem, left/right cochlea, left/right eye, larynx, left/right lens, mandible, optic chiasm, left/right optic nerve, oral cavity, left/right parotid, pharynx, and spinal cord, respectively, were achieved.  
	
	\textbf{Conclusion:} In this study, we developed a synthetic MRI-aided HN CBCT auto-segmentation method based on deep learning. It provides a rapid and accurate OAR auto-delineation approach, which can be used for adaptive radiation therapy.
	
	\bigbreak
	\bigbreak
	
	\textbf{keywords:} Image segmentation, cone-beam CT, adaptive radiation therapy, synthetic MRI, radiotherapy, deep learning.

	\noindent 
	\section{ INTRODUCTION}
	
	Head and neck (HN) cancers involving regions like oral cavity, pharynx, larynx, salivary glands, paranasal sinuses, and nasal cavity account for about 4\% of all cancers in the United States\cite{RN1741}. Radiotherapy, along with surgery and chemotherapy, is commonly used for treatment of HN cancers\cite{RN1740}. HN tumors are generally very close to organs-at-risk (OARs), thus, steep dose gradients are highly desired in radiotherapy to deliver therapeutic doses to the targets while better sparing the OARs. Modern radiotherapy techniques including intensity-modulated radiation therapy (IMRT)\cite{RN1745}, volumetric-modulated arc therapy (VMAT)\cite{RN1746}, and intensity-modulated proton therapy (IMPT)\cite{RN1747} offer highly conformal dose delivery to target regions meanwhile sparing adjacent OARs. However, these techniques often show higher sensitivity to geometrical changes resulting from uncertainties in patient setup and inter-fractional changes in patient anatomy, for example, the planned dose distribution in IMPT can be distorted by these changes that are not recorded in the planning CT, which impacting the quality of treatment\cite{RN1744, RN1743, RN1742}. Generally, certain margins around the target are created taking into account these uncertainties, to guarantee the quality of treatment, which essentially results in increasing toxicity in OARs\cite{RN1795}. Wang et al.\cite{RN1796} reported that with the online correction using daily CBCT, the planning target volume (PTV) margins can be reduced to 3 mm from 5-6 mm without correction in IMRT for nasopharyngeal tumors. They further show that with plan adaptation, the maximum dose to both brain stem and spinal cord could be reduced by 10 Gy, and the mean dose to the left/right parotid glands could also be reduced by approximately 8 Gy.
	
	Adaptive radiation therapy\cite{RN1749, RN1748} has been proposed to deal with uncertainties to target dose coverage while minimizing dosage to OARs. Offline treatment re-planning based on re-scan CT when the original plan cannot fulfill clinical goals is a common strategy in current clinical practice\cite{RN1756, RN1755}. Online plan adaptation has been proposed as a promising alternative, through which, treatment delivery will not be delayed, the adaptation takes into account concurrent setup and patient anatomy, and more importantly, margins can be potentially reduced\cite{RN1760, RN1758}. Online adaptive radiation therapy requires online imaging system which can offer patient anatomy and contours of target and OARs at treatment position for re-planning process that should be completed in a few minutes in order to fit into current clinical workflow\cite{RN1770}. CBCT is a widely used in-room imaging for patient positioning in modern radiotherapy including IMRT, VMAT, and IMPT, which shows high potentials for online plan adaptation\cite{RN1762, RN1758}. Accurately delineating tumor targets and OARs is an important step in adaptive re-planning process\cite{RN1862, RN1861}, however, manual delineation can be labor-intensive and time-consuming, especially, for HN cancer cases, it could highly prolong the entire adaptational procedure due to the multiple OARs have to be contoured. Therefore, fully automated delineation method with a high quality of performance is highly desired to expedite the contouring process of adaptive radiotherapy re-planning and dose-volume based plan evaluation and monitoring. However, simultaneous delineation of multiple OARs on CBCT is a challenging task due to the relatively poor image quality, low soft-tissue contrast, and high image noise resulting from scattering compared to planning CT\cite{RN81, RN1411}.
	
	CBCT-based OAR delineation can be grouped into two main classes: (1) deformation of contours in the original planning CT based on CBCT information, and (2) direct CBCT-based delineation. Using deformable image registration algorithms, the contours in the original planning CT can be transferred to CBCT\cite{RN1764, RN1753}. Veiga et al.\cite{RN1755} studied B-spline free form deformation algorithm based registration from planning CT to a daily CBCT for calculation of daily dosage without re-acquiring a new CT, which is a first step of implementation of online plan adaptation. Landry et al.\cite{RN1766} investigated deformable image registration-based approach for translating from CBCT and planning CT to a virtual CT which is then used for proton dose recalculation. The recalculated dose distributions on the virtual CT agreed well to those calculated on the re-planning. Kurz et al.\cite{RN1758} further investigated plan adaptation in IMPT of HN cancers using delineations from deformable registration of the planning CT to CBCT. Botas et al.\cite{RN1760} studied online IMPT plan adaptation for HN cancer patients based on CBCT, where the contours were propagated to CBCT with a vector filed computed with deformable registration between planning CT and CBCT. In plan adaptation, deformable image registration enables transferring contours in original planning CT to CBCT accounting for anatomic changes. Commercial deformable image registration platforms are available in clinical practice for propagating contours with CBCT such as Velocity (Varian Medical Systems), RayStation (RaySearch Laboratories), and MIM Maestro (MIM Software Inc.)\cite{RN1771, RN1772, RN1773}.  However, the performance of deformable image registration varies with algorithm, implementation, and image quality\cite{RN1774}, which will bring uncertainties to the contours propagation from CT to CBCT, eventually adding uncertainties in dose evaluation\cite{RN1756}. Proper selection of a specific deformable image registration method which depends on image modality, anatomic site, and the magnitude of deformation is required for obtaining acceptable propagated contours for plan adaptation\cite{RN1767}.
	
	Direct CBCT-based delineation overcomes the propagational uncertainties from registration errors between planning CT and CBCT. Caucig et al.\cite{RN48} investigated the feasibility of delineating targets and OARs using CBCT in prostate radiotherapy. While inspiring, their results show greater interobserver variability in delineations for CBCT-based contouring than CT or MRI-based. Patient-specific model-aided CBCT segmentation methods were proposed to improve the performance of delineation\cite{RN5, RN55}. Altorjai et al.\cite{RN1775} studied CBCT-based delineation of lung lesions for adaptive stereotactic body radiotherapy, and no significant differences in interobserver variability was seen in CBCT-based stereotactic lung target delineation compared to CT-based. Very recently, our group developed a synthetic MRI aided CBCT multi-organ segmentation method for prostate radiotherapy\cite{RN7}. In this study, synthetic MRI (sMRI) was firstly generated given CBCT images using a cycle-consistent generative adversarial network (cycleGAN) trained by CBCT and MRI pairs, and was then paired with their corresponding manual contours for training a neural network for sMRI segmentation. The results show encouraging accuracy in delineation of prostate, bladder and rectum compared to physicians’ manual contours. 
	
	In this study, we adopt the synthetic-MRI-aided strategy from our previous work, further improve the network architecture using dual pyramid networks, and investigate the feasibility of automatic delineating OARs in HN cancer radiation therapy. Again, a pre-trained cycleGAN is firstly used to synthesize MR images given the CBCT images, and the features of CBCT and synthetic MRI images are then extracted using dual pyramid networks for final delineation of organs. Manually verified contours from CBCT were used as ground truth contours for training and evaluation of the proposed method. The proposed method aims to offer accurate OAR contours using only single CBCT modality, which greatly accelerates the re-planning process thus facilitates online plan adaptation for modern radiation therapy.

	\noindent 
	\section{Methods and Materials}
	\noindent 
	\subsection{Method overview}
	
	CycleGAN has achieved encouraging performance in image-to-image translation when the paired images are absent\cite{RN1609, RN1607, RN3815, RN3838}. CycleGAN learns a mapping G:X→Y such that the distribution of images from G(X) is indistinguishable from the distribution Y using an adversarial loss. Cycle consistent loss force the mapping G to be closed to a one-to-one mapping. In medical imaging, cycleGAN has demonstrated as a unique image synthesis method. As shown in Figure 1, a cycleGAN model was used to synthesize MRI taking in CBCT images as input. Feature pyramid network (FPN) was original proposed to improve the performance of deep convolutional networks-based algorithms for object detection through collecting pyramid representations which are basic components in recognition systems at different scales\cite{RN1776}. To avoid memory intensive computations, FPN strategically exploits the inherent multi-scale and pyramidal hierarchy of deep convolutional networks by combining bottom-up and top-down pathways. Two separated FPN models were used for extracting features from CBCT and sMRI, respectively. These extracted multi-scale feature maps were then combined through attention gating for final segmentation. We named this novel architecture as dual pyramid networks (DPN), which will be introduced in detail in the following section.

	\begin{figure}
		\centering
		\noindent \includegraphics*[width=6.50in, height=4.20in, keepaspectratio=true]{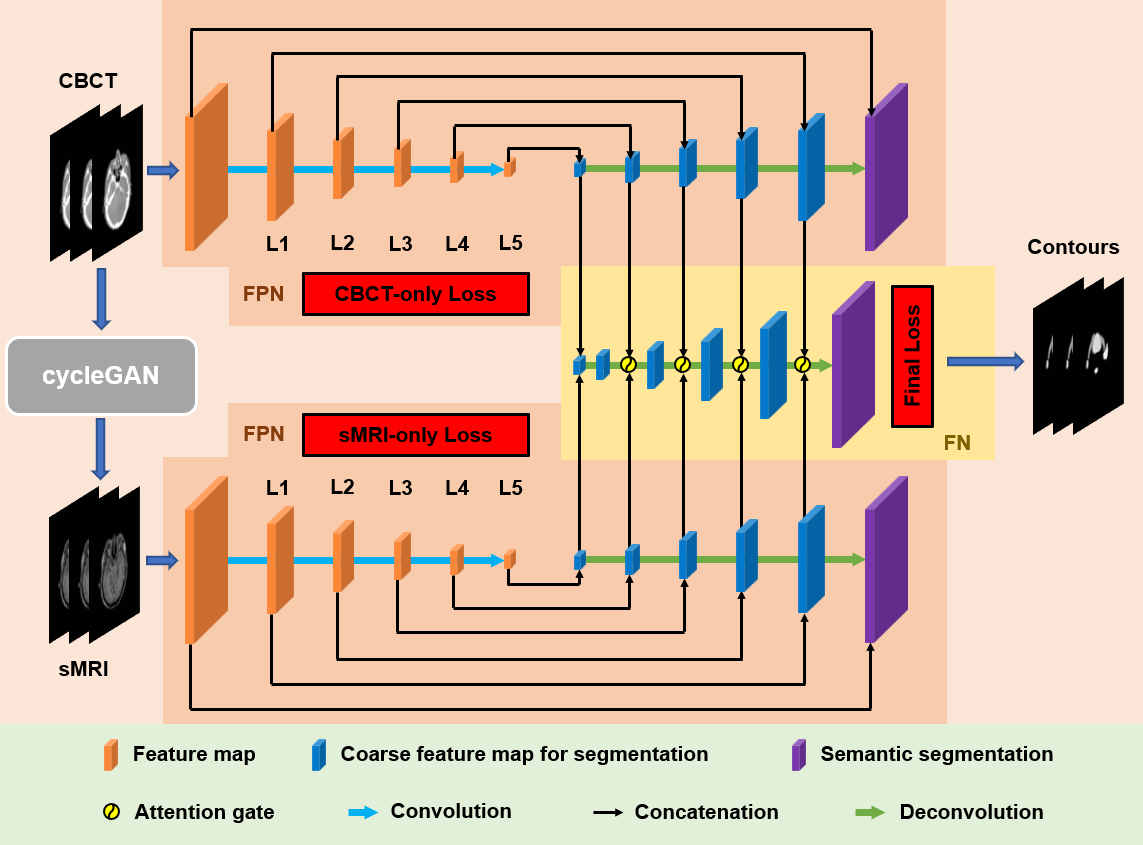}
		
		\noindent Fig. 1. The schematic flow diagram of the proposed method for multi-organ HN OAR auto-delineation. sMRI, synthetic MRI; cycleGAN, cycle-consistent generative adversarial network; L1 – L5, hierarchic resolution levels for feature map; FPN, feature pyramid network; FN, fusion network.
	\end{figure}

	\noindent 
	\subsection{MRI synthesis}
	
	The details of cycleGAN can be found in our previous study\cite{RN7}. Figure 2 briefly shows the network architecture of the cycleGAN for synthesis of MRI from CBCT. The generator was constructed with one convolutional layer with stride size of 1×1×1, two convolutional layers with stride size of 2×2×2,  nine ResNet blocks, two deconvolution layers and two convolutional layers with stride size of 1×1×1. The discriminator consisted of six convolutional layers with stride size of 2×2×2, followed by one fully connected layer and a sigmoid operation.
	
	\begin{figure}
		\centering
		\noindent \includegraphics*[width=6.50in, height=4.20in, keepaspectratio=true]{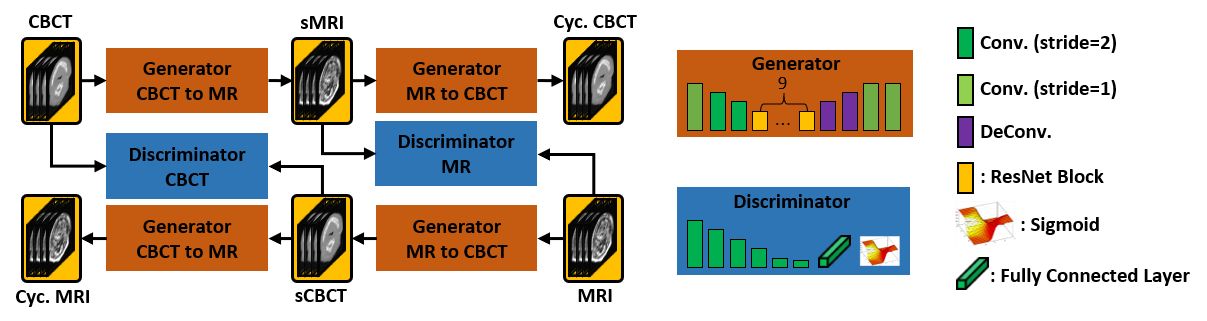}
		
		\noindent Fig. 2. The network architecture of cycleGAN for MRI synthesis.
	\end{figure}

	\noindent 
	\subsection{Dual pyramid networks}
	
	As shown in Figure 1, in our proposed dual pyramid networks, two feature pyramid networks (FPNs) are incorporated in parallel to extract features from CBCT and sMRI independently. Adopting the architecture from the work of Lin et al.\cite{RN1776}, each FPN is constructed of a bottom-up pathway, a top-down pathway, and lateral connections. The bottom-up pathway is the forward convolutional operations that computes hierarchical feature maps at several scales. One feature pyramid level (e.g., L1 - L5, in Figure 1) is defined at each scale. The top-down pathway is composed by deconvolutions, where, higher resolution features are hallucinated by up-sampling from higher pyramid levels that are spatially coarser but semantically stronger. Those features in the top-down pathway are enhanced by the lateral connections which combine the feature maps from the bottom-up and the top-down pathways that have the same spatial resolutions. Aiming to force the feature maps extracted from the two FPNs to be mostly correlated to the semantic segmentation of target organs, two separated compound segmentation loss functions, namely, CBCT-only loss and sMRI-only loss, are applied to supervise the CBCT and sMRI FPNs, respectively. These feature maps from the two FPNs at multiple scales are then combined to a fusion network (FN). Inspired by the attention gated networks\cite{RN1675} which can highlight salient features of images and preserve only relevant activations to a specific task such as segmentation, the fusion network is constructed as an attention gated network, in which, the features extracted from CBCT and sMRI at the same pyramid level in the FPNs are firstly concatenated. The concatenated maps are then up-sampled to hallucinate the feature maps in a higher resolution, which are combined with the feature maps from both CBCT and sMRI through attention gating. The detailed implementation of the attention gates can refer to recent studies\cite{RN1675}. Shortly, a concatenated tensor is firstly computed through linear transformations using channel-wise 1x1x1 convolutions from input feature maps, sequentially followed by ReLU, 1x1x1 convolution, and sigmoid operations to compute the attention coefficients. The final attention gates are calculated by element-wise multiplication of input feature maps and the attention coefficients. The final contours of multiple organs can be obtained from the output of the fusion network which is constructed as a top-down structure. Again, an independent compound segmentation loss (denoted as final loss in Figure 1) is applied to supervise the learnable parameters of the fusion network.
	
	The compound segmentation loss function used in our proposed method is define as a combination of Dice loss and binary cross entropy (BCE) in deep supervision manner, which has been widely used for medical image segmentation, can be expressed as:
	\begin{equation} 
	L_{compunded}\left(C,\hat{C}\right)=\sum_{l=1}^{N}{\lambda_l\left(L_{BCE}\left(C,{\hat{C}}^l\right)+\mu L_{DSC}\left(C,{\hat{C}}^l\right)\right)}
	\end{equation} 
	where $C$ and $\hat{C}$ are the ground truth contour and prediction, respectively, $l$ denotes the pyramid stage of the network, $N$ is the total number of the stages in the network, $\lambda_l$ is the weight of the lth pyramid stage’s loss, and $\mu$ is the parameter that balances the BCE loss $L_{BCE}\left(C,\hat{C}\right)$ and Dice loss.   
	
	The BCE loss is defined as: 
	\begin{equation} 
	L_{BCE}\left(C,\hat{C}\right)=-\sum_{j}\left[C_jlog{{\hat{C}}_j}+\left(1-C_j\right)log{\left(1-{\hat{C}}_j\right)}\right]                    
	\end{equation} 
	where, $C_j$ and ${\hat{C}}_j$ denote the jth voxel in ground truth contour $C$ and prediction $\hat{C}$, respectively.
	
	The Dice loss can be expressed as:
	\begin{equation} 
	L_{DSC}\left(C,\hat{C}\right)=1-\frac{2V\left(C\cap\hat{C}\right)}{V\left(C\right)+V\left(\hat{C}\right)}                     
	\end{equation}
	where,$V$ indicates the volume of the region enclosed in the contours.
	
	\noindent 
	\subsection{Image data acquisition, implementation and experimental setting}
	
	A retrospective study was conducted on a cohort of 65 HN cancer patients receiving proton therapy. Each dataset includes a planning CT, MRI, one CBCT acquired during treatment, and a set of multi-organ contours. The MRIs were acquired by a Siemens Aera 1.5T scanner as T1-weigted DIXON opposed-phase maps. CBCTs were acquired by an in-room CBCT scanner on a Varian ProBeam system. For validation of our proposed method, these multi-modality images were reconstructed and re-sampled to have a uniformed voxel size of 0.9766 mm × 0.9766 mm × 1.0 mm using commercial medical image processing software (Velocity AI 3.2.1, Varian Medical Systems, Palo Alto, CA). Considering the tumor shrinkage and deformation after radiation, we only selected the CBCT which was captured in the first fraction before radiation. Our MRI and CT were captured in the same simulation day. Usually, the time gap between simulation and first treatment is around 1 week in our clinic. In this study we assumed that the difference between MRI and CBCT (the first fraction) is minor and ignore. The manual contours were firstly made by radiation oncologists based on CT and MRI, and then propagated to CBCT through deformable registration using Velocity AI 3.2.1. The propagated contours on CBCT were then adjusted and confirmed by the radiation oncologists to obtain the final multi-organ delineations, which were utilized as ground truth, pairing with CBCT to train and test our proposed method. Therefore, the errors from registration and propagation were finally eliminated by physicians’ revisions.
	
	The proposed method was implemented using python 3.6.8 and Tensorflow 1.8.0. The cycleGAN was pre-trained on a NVIDIA Tesla V100 GPU with 32 GB of memory. sMRI can be generated in approximated 2-5 minutes after the model was well-trained. In our previous work\cite{RN49}, we discussed the detailed information about the implementation of cycleGAN. The dual pyramid networks were trained using Adam optimizer with a learning rate of 2e-4 for 400 epochs. For each epoch, the batch size was set to 20 to fit the 32 GB memory of the NVIDIA Tesla V100 GPU. The training took around 6 hours, while the multi-organ contours can be predicted in about 1 second. For the experiments, hold-out validation was performed, where, thirty-five patients were randomly selected as the training data, and the rest thirty patients were used as hold-out test data.

	\noindent 
	\subsection{Evaluation metrics}
	
	The performance of our proposed method for multiple OARs automated contouring were quantitatively assessed by computing quantities including Dice similarity coefficient (DSC), 95th percentile Hausdorff distance (HD95), mean surface distance (MSD), and residual mean square distance (RMS). DSC is used to gauge the similarity of two samples. It is widely used in medical image segmentation, particularly for comparing algorithm output against reference masks\cite{RN1778, RN1777}. DSC can be expressed as:
	
	\begin{equation}
	DSC=\frac{2\left|X\cap Y\right|}{\left|X\right|+\left|Y\right|}
	\end{equation}
	where,$\left|X\right|$ and $\left|Y\right|$ are the cardinalities of the ground truth mask X and predicted mask Y, respectively.
	 
	Hausdorff distance, MSD, and RMS are three quantities that measure the surface organ distance between the ground truth and the predicted contours. HD95 is can be expressed as\cite{RN1780, RN1779}:
	\begin{equation}
	HD95=max\left[{\vec{d}}_{H,95}\left(X,Y\right),\ \ {\vec{d}}_{H,95}\left(Y,X\right)\right]
	\end{equation}
	where,${\vec{d}}_{H,95}\left(X,Y\right)$ is 95\% percentile Hausdorff distance which measures the 95th percentile distance of all distances between points in X and the nearest point in Y, defined as $ {\vec{d}}_{H,95}\left(X,Y\right)=\ K_{95}({min}_{y\in\left|Y\right|}d(x,y))$ , and $d(x,y)$ is Euclidean distance.

	MSD is defined as:
	\begin{equation}
	MSD=\frac{1}{\left|X\right|+\left|Y\right|}\left(\sum_{x\epsilon X}{\vec{d}\left(x,Y\right)+\sum_{y\epsilon Y}{\vec{d}\left(y,X\right)}}\right)
	\end{equation}
	where, $\vec{d}\left(x,Y\right)=\ {min}_{y\in\left|Y\right|}d(x,y)$
	
	RMS can be expressed as:	
	\begin{equation}
	 RMS=\ \sqrt{\frac{1}{\left|X\right|+\left|Y\right|}\left(\sum_{x\epsilon\left|X\right|}{{\vec{d}\left(x,Y\right)}^2+\sum_{y\epsilon\left|Y\right|}{\vec{d}\left(y,X\right)}^2}\right)}
	\end{equation}
	
	In general, the closer the DSC value to 1, the smaller HD95, MSD, and RMS values, the higher similarity between predicted and ground truth contours, therefore, the better the performance of the segmentation algorithm.
	
	\noindent 
	\section{Results}
	\subsection{Overall performance}
	
	For comparison, we conducted all the experiments on both the proposed method (CBCT+sMRI), the approach using the same network (dual pyramid networks) excluding cycleGAN but only taking CBCT as inputs (two channels of CBCTs were used as inputs of the DPN without synthetic MRI), which we named as CBCT\_only, and the method using dual pyramid networks but taking sMRI for the two channel inputs, which we named as sMRI\_only approach. Figure 3 shows the results from one patient. Five different transverse slices are shown in five rows. The leftmost two columns are respectively CBCT and sMRI images, followed by the next four columns respectively showing the ground truth, the discrepancies of the proposed method predicted, CBCT\_only predicted, sMRI\_only predicted contours referring to the ground truth. Figure 4 shows the three orthogonal views (axial, sagittal, and coronal) of another case. The rows from left to right show the CBCT, sMRI, the ground truth, the differences between the ground truth and the predictions by our proposed, CBCT\_only, sMRI\_only methods respectively. Qualitatively investigating on the images in Figures 3 and 4, we can see, the predicted contours from our proposed method have less discrepancies with the ground truth than either CBCT\_only or sMRI\_only method, attribute to the strength of combining the complementary features extracted from CBCT and synthetic MRI images. 
	
	\begin{figure}
		\centering		
		\noindent \includegraphics*[width=6.50in, height=4.20in, keepaspectratio=true]{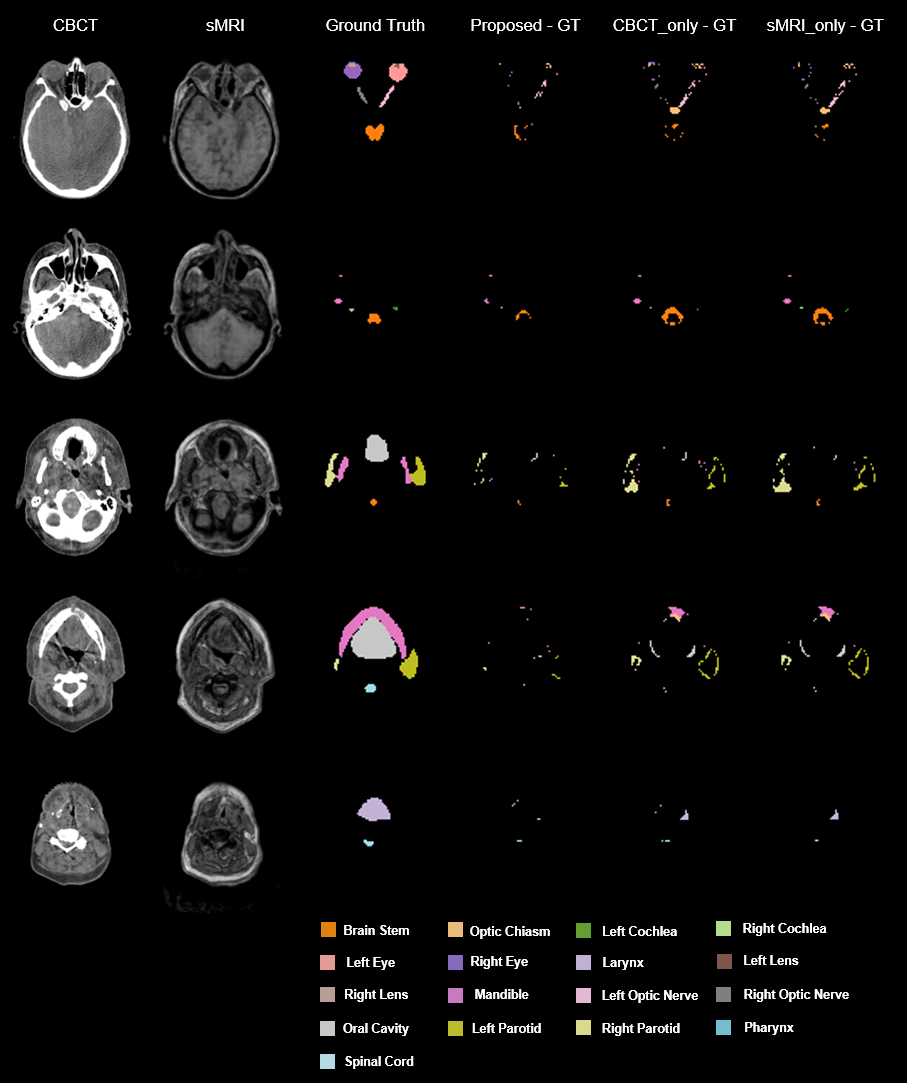}
		
		\noindent Fig. 3. An illustrative example of the benefit of our proposed method compared with CBCT\_only, and sMRI\_only methods. sMRI, synthetic MRI; Proposed – GT, the discrepancy between the proposed predicted contours and the ground truth; CBCT\_only – GT, the difference between CBCT\_only predicted contours and the ground truth; sMRI\_only – GT, the discrepancy between sMRI\_only prediction and the ground truth.
	\end{figure}

	\begin{figure}
		\centering		
		\noindent \includegraphics*[width=6.50in, height=4.20in, keepaspectratio=true]{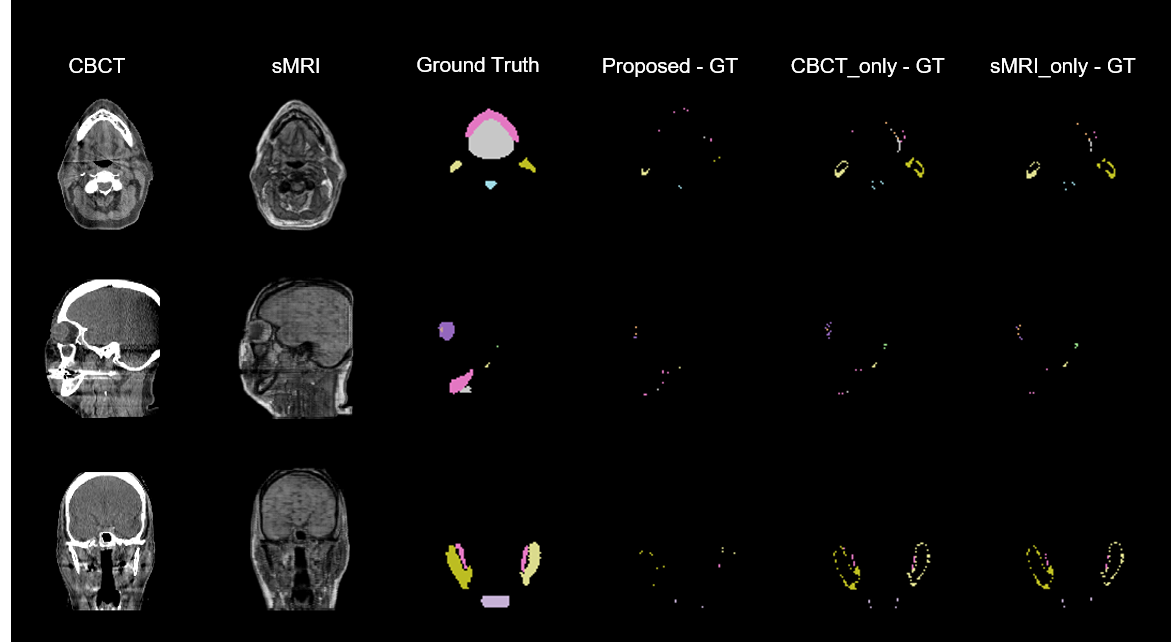}
		
		\noindent Fig. 4. Three orthogonal views of one case. sMRI, synthetic MRI; Proposed – GT, the discrepancy between the proposed predicted contours and the ground truth; CBCT\_only – GT, the difference between CBCT\_only predicted contours and the ground truth; sMRI\_only – GT, the discrepancy between sMRI\_only prediction and the ground truth. 
	\end{figure}
	
	\subsection{Quantitative results}
	
	To quantitatively show the improvements of our synthetic MRI-aided CBCT segmentation method, metrics including DSC, HD95, MSD, and RMS were calculated. The results of CBCT\_only, sMRI\_only and our proposed methods were summarized in Table 1, where, the mean and standard deviation of these metrics are listed for organs including brain stem, left/right cochlea, left/right eye, larynx, left/right lens, mandible, optic chiasm, left/right optic nerve, oral cavity, left/right parotid, pharynx, and spinal cord that are commonly contoured in radiotherapy treatment planning. Our proposed method achieved the DSC values of as high as 0.87 ± 0.03, 0.79 ± 0.10 / 0.79 ± 0.11, 0.89 ± 0.08 / 0.89 ± 0.07, 0.90 ± 0.08, 0.75 ± 0.06 / 0.77 ± 0.06, 0.86 ± 0.13, 0.66 ± 0.14, 0.78 ± 0.05 / 0.77 ± 0.04, 0.96 ± 0.04, 0.89 ± 0.04 / 0.89 ± 0.04, 0.83 ± 0.02, and 0.84 ± 0.07 for brain stem, left/right cochlea, left/right eye, larynx, left/right lens, mandible, optic chiasm, left/right optic nerve, oral cavity, left/right parotid, pharynx, and spinal cord, respectively. Furthermore, to analyze the statistical significance of the improvement by our proposed method compared to CBCT\_only and sMRI\_only methods, paired student t-test was conducted for all the metrics, the results of which are shown in Table 2. From Tables 1 and 2, overall, the CBCT+sMRI method outperforms either CBCT\_only or sMRI\_only method in all the four metrics including DSC, HD95, MSD, and RMS. By closely investigating on the DSC comparison, our proposed method achieved significant improvements (p < 0.05) in organs including left/right cochlea, left/right eye, left/right lens, mandible, left/right optic nerve, oral cavity, and left/right parotid compared to CBCT\_only method, and in organs including left/right optic nerve, oral cavity, and left/right parotid compared to sMRI\_only method.
	
	\begin{table}[htbp]
		\centering
		\caption{Overall quantitative results achieved by the CBCT\_only, sMRI\_only, and the CBCT+sMRI methods.}
		\begin{tabular}{llllll}
			\hline
			Organ                                & Method    & DSC       & HD95 (mm) & MSD (mm)  & RMS (mm)  \\ \hline
			\multirow{3}{*}{Brain   Stem}        & CBCT-only & 0.86±0.10 & 4.91±4.13 & 1.37±1.13 & 2.17±1.70 \\ \cline{2-6} 
			& sMRI-only & 0.87±0.09 & 4.20±3.25 & 1.26±1.00 & 1.91±1.38 \\ \cline{2-6} 
			& CBCT+sMRI & 0.87±0.03 & 2.96±0.93 & 1.10±0.31 & 1.49±0.43 \\ \hline
			\multirow{3}{*}{Left   Cochlea}      & CBCT-only & 0.74±0.10 & 1.52±0.22 & 0.77±0.21 & 0.70±0.34 \\ \cline{2-6} 
			& sMRI-only & 0.76±0.09 & 1.29±0.33 & 0.47±0.14 & 0.73±0.13 \\ \cline{2-6} 
			& CBCT+sMRI & 0.79±0.10 & 1.15±0.31 & 0.43±0.19 & 0.67±0.18 \\ \hline
			\multirow{3}{*}{Right   Cochlea}     & CBCT-only & 0.75±0.13 & 1.35±0.50 & 0.51±0.28 & 0.76±0.28 \\ \cline{2-6} 
			& sMRI-only & 0.75±0.11 & 1.35±0.38 & 0.50±0.22 & 0.76±0.24 \\ \cline{2-6} 
			& CBCT+sMRI & 0.79±0.11 & 1.17±0.40 & 0.44±0.21 & 0.68±0.21 \\ \hline
			\multirow{3}{*}{Left   Eye}          & CBCT-only & 0.83±0.15 & 3.66±3.54 & 1.16±1.08 & 1.68±1.52 \\ \cline{2-6} 
			& sMRI-only & 0.88±0.12 & 1.88±1.75 & 0.70±0.48 & 1.03±0.74 \\ \cline{2-6} 
			& CBCT+sMRI & 0.89±0.08 & 1.88±1.20 & 0.68±0.36 & 1.01±0.51 \\ \hline
			\multirow{3}{*}{Right   Eye}         & CBCT-only & 0.84±0.13 & 3.34±2.66 & 1.06±0.78 & 1.56±1.09 \\ \cline{2-6} 
			& sMRI-only & 0.87±0.14 & 1.86±1.76 & 0.72±0.41 & 1.02±0.60 \\ \cline{2-6} 
			& CBCT+sMRI & 0.89±0.07 & 1.74±0.86 & 0.69±0.35 & 0.97±0.40 \\ \hline
			\multirow{3}{*}{Larynx}              & CBCT-only & 0.88±0.19 & 5.82±7.85 & 1.56±2.76 & 2.54±3.90 \\ \cline{2-6} 
			& sMRI-only & 0.85±0.20 & 5.80±6.39 & 1.67±2.10 & 2.60±2.95 \\ \cline{2-6} 
			& CBCT+sMRI & 0.90±0.08 & 2.64±1.74 & 0.90±0.54 & 1.33±0.74 \\ \hline
			\multirow{3}{*}{Left   Lens}         & CBCT-only & 0.64±0.15 & 1.99±0.67 & 0.74±0.30 & 1.04±0.33 \\ \cline{2-6} 
			& sMRI-only & 0.71±0.09 & 1.52±0.31 & 0.66±0.10 & 0.92±0.09 \\ \cline{2-6} 
			& CBCT+sMRI & 0.75±0.06 & 1.45±0.40 & 0.58±0.16 & 0.83±0.17 \\ \hline
			\multirow{3}{*}{Right   Lens}        & CBCT-only & 0.66±0.07 & 2.32±1.14 & 0.91±0.39 & 1.24±0.51 \\ \cline{2-6} 
			& sMRI-only & 0.71±0.09 & 1.36±0.34 & 0.61±0.14 & 0.85±0.17 \\ \cline{2-6} 
			& CBCT+sMRI & 0.77±0.06 & 1.28±0.23 & 0.47±0.08 & 0.72±0.09 \\ \hline
			\multirow{3}{*}{Mandible}            & CBCT-only & 0.81±0.13 & 4.38±6.34 & 1.31±1.06 & 2.22±2.29 \\ \cline{2-6} 
			& sMRI-only & 0.85±0.13 & 3.97±6.00 & 1.02±0.98 & 1.92±2.11 \\ \cline{2-6} 
			& CBCT+sMRI & 0.86±0.13 & 2.13±1.86 & 0.89±0.94 & 1.48±1.68 \\ \hline
			\multirow{3}{*}{Optic   Chiasm}      & CBCT-only & 0.65±0.16 & 3.82±2.53 & 1.02±0.71 & 1.71±1.09 \\ \cline{2-6} 
			& sMRI-only & 0.62±0.15 & 3.33±1.47 & 0.95±0.46 & 1.51±0.63 \\ \cline{2-6} 
			& CBCT+sMRI & 0.66±0.14 & 2.69±2.36 & 0.80±0.55 & 1.28±0.94 \\ \hline
			\multirow{3}{*}{Left   Optic Nerve}  & CBCT-only & 0.68±0.09 & 6.36±7.59 & 1.57±2.11 & 2.75±3.40 \\ \cline{2-6} 
			& sMRI-only & 0.70±0.08 & 5.71±6.33 & 1.20±1.09 & 2.35±2.25 \\ \cline{2-6} 
			& CBCT+sMRI & 0.78±0.05 & 1.86±1.73 & 0.55±0.18 & 0.93±0.47 \\ \hline
			\multirow{3}{*}{Right   Optic Nerve} & CBCT-only & 0.68±0.09 & 6.59±6.25 & 1.31±0.99 & 2.58±2.10 \\ \cline{2-6} 
			& sMRI-only & 0.70±0.07 & 5.32±6.02 & 0.87±0.47 & 1.62±1.07 \\ \cline{2-6} 
			& CBCT+sMRI & 0.77±0.04 & 2.06±2.69 & 0.58±0.32 & 1.04±0.85 \\ \hline
			\multirow{3}{*}{Oral   Cavity}       & CBCT-only & 0.90±0.10 & 4.29±4.49 & 1.58±1.61 & 2.14±2.13 \\ \cline{2-6} 
			& sMRI-only & 0.92±0.10 & 4.03±4.29 & 1.33±1.62 & 1.94±2.08 \\ \cline{2-6} 
			& CBCT+sMRI & 0.96±0.04 & 2.98±1.55 & 0.68±0.54 & 1.29±0.72 \\ \hline
			\multirow{3}{*}{Left   Parotid}      & CBCT-only & 0.86±0.03 & 3.69±0.96 & 1.28±0.30 & 1.90±0.62 \\ \cline{2-6} 
			& sMRI-only & 0.87±0.02 & 2.78±0.66 & 1.08±0.20 & 1.46±0.32 \\ \cline{2-6} 
			& CBCT+sMRI & 0.89±0.04 & 2.39±0.86 & 0.95±0.32 & 1.29±0.39 \\ \hline
			\multirow{3}{*}{Right   Parotid}     & CBCT-only & 0.85±0.07 & 3.71±0.96 & 1.32±0.45 & 1.85±0.59 \\ \cline{2-6} 
			& sMRI-only & 0.87±0.03 & 2.64±0.56 & 1.10±0.19 & 1.48±0.27 \\ \cline{2-6} 
			& CBCT+sMRI & 0.89±0.04 & 2.26±0.71 & 0.88±0.24 & 1.18±0.31 \\ \hline
			\multirow{3}{*}{Pharynx}             & CBCT-only & 0.75±0.14 & 4.36±3.55 & 1.10±0.87 & 1.91±1.40 \\ \cline{2-6} 
			& sMRI-only & 0.83±0.04 & 1.95±0.57 & 0.59±0.11 & 0.97±0.21 \\ \cline{2-6} 
			& CBCT+sMRI & 0.83±0.02 & 1.45±0.26 & 0.58±0.08 & 0.86±0.07 \\ \hline
			\multirow{3}{*}{Spinal   Cord}       & CBCT-only & 0.83±0.07 & 6.48±7.75 & 1.27±1.00 & 2.77±2.72 \\ \cline{2-6} 
			& sMRI-only & 0.82±0.10 & 7.30±9.20 & 1.46±1.31 & 3.08±3.06 \\ \cline{2-6} 
			& CBCT+sMRI & 0.84±0.07 & 2.28±1.01 & 0.91±0.40 & 1.21±0.48 \\ \hline
		\end{tabular}
	\end{table}

	\begin{table}[htbp]
		\centering
		\caption{Results of paired student t-test (p-value) between the CBCT\_only and the CBCT+sMRI, the sMRI\_only and the CBCT+sMRI methods.}
		\begin{tabular}{llllll}
			\hline
			Organ                                & Method                   & DSC             & HD95            & MSD             & RMS             \\ \hline
			\multirow{2}{*}{Brain   Stem}        & CBCT-only   vs. proposed & 0.54            & 0.03            & 0.24            & 0.06            \\ \cline{2-6} 
			& sMRI-only   vs. proposed & 0.66            & 0.06            & 0.42            & 0.13            \\ \hline
			\multirow{2}{*}{Left   Cochlea}      & CBCT-only   vs. proposed & \textless{}0.01 & \textless{}0.01 & \textless{}0.01 & 0.03            \\ \cline{2-6} 
			& sMRI-only   vs. proposed & 0.23            & 0.11            & 0.17            & 0.12            \\ \hline
			\multirow{2}{*}{Right   Cochlea}     & CBCT-only   vs. proposed & 0.02            & 0..02           & 0.03            & 0.03            \\ \cline{2-6} 
			& sMRI-only   vs. proposed & 0.31            & 0.11            & 0.28            & 0.20            \\ \hline
			\multirow{2}{*}{Left   Eye}          & CBCT-only   vs. proposed & 0.02            & \textless{}0.01 & 002             & 0.02            \\ \cline{2-6} 
			& sMRI-only   vs. proposed & 0.82            & 0.99            & 0.91            & 0.93            \\ \hline
			\multirow{2}{*}{Right   Eye}         & CBCT-only   vs. proposed & 0.02            & \textless{}0.01 & 0.02            & 0.02            \\ \cline{2-6} 
			& sMRI-only   vs. proposed & 0.52            & 0.81            & 0.81            & 0.79            \\ \hline
			\multirow{2}{*}{Larynx}              & CBCT-only   vs. proposed & 0.62            & 0.07            & 0.28            & 0.17            \\ \cline{2-6} 
			& sMRI-only   vs. proposed & 0.07            & 0.01            & 0.06            & 0.03            \\ \hline
			\multirow{2}{*}{Left   Lens}         & CBCT-only   vs. proposed & 0.04            & 0.03            & 0.04            & 0.04            \\ \cline{2-6} 
			& sMRI-only   vs. proposed & 0.37            & 0.50            & 0.12            & 0.20            \\ \hline
			\multirow{2}{*}{Right   Lens}        & CBCT-only   vs. proposed & \textless{}0.01 & 0.03            & 0.01            & 0.02            \\ \cline{2-6} 
			& sMRI-only   vs. proposed & 0.13            & 0.41            & 0.06            & 0.11            \\ \hline
			\multirow{2}{*}{Mandible}            & CBCT-only   vs. proposed & \textless{}0.01 & 0.23            & 0.11            & 0.25            \\ \cline{2-6} 
			& sMRI-only   vs. proposed & 0.66            & 0.29            & 0.60            & 0.50            \\ \hline
			\multirow{2}{*}{Optic   Chiasm}      & CBCT-only   vs. proposed & 0.86            & 0.15            & 0.26            & 0.19            \\ \cline{2-6} 
			& sMRI-only   vs. proposed & 0.47            & 0.37            & 0.37            & 0.42            \\ \hline
			\multirow{2}{*}{Left   Optic Nerve}  & CBCT-only   vs. proposed & \textless{}0.01 & 0.04            & 0.04            & 0.04            \\ \cline{2-6} 
			& sMRI-only   vs. proposed & 0.01            & 0.04            & 0.04            & 0.04            \\ \hline
			\multirow{2}{*}{Right   Optic Nerve} & CBCT-only   vs. proposed & \textless{}0.01 & \textless{}0.01 & \textless{}0.01 & \textless{}0.01 \\ \cline{2-6} 
			& sMRI-only   vs. proposed & \textless{}0.01 & 0.04            & 0.02            & 0.04            \\ \hline
			\multirow{2}{*}{Oral   Cavity}       & CBCT-only   vs. proposed & \textless{}0.01 & 0.04            & \textless{}0.01 & 0.04            \\ \cline{2-6} 
			& sMRI-only   vs. proposed & 0.03            & 0.02            & 0.04            & 0.04            \\ \hline
			\multirow{2}{*}{Left   Parotid}      & CBCT-only   vs. proposed & 0.02            & \textless{}0.01 & \textless{}0.01 & \textless{}0.01 \\ \cline{2-6} 
			& sMRI-only   vs. proposed & 0.01            & 0.03            & 0.03            & 0.03            \\ \hline
			\multirow{2}{*}{Right   Parotid}     & CBCT-only   vs. proposed & 0.02            & \textless{}0.01 & \textless{}0.01 & \textless{}0.01 \\ \cline{2-6} 
			& sMRI-only   vs. proposed & \textless{}0.01 & \textless{}0.01 & \textless{}0.01 & \textless{}0.01 \\ \hline
			\multirow{2}{*}{Pharynx}             & CBCT-only   vs. proposed & 0.13            & 0.04            & 0.12            & 0.05            \\ \cline{2-6} 
			& sMRI-only   vs. proposed & 0.78            & 0.06            & 0.86            & 0.18            \\ \hline
			\multirow{2}{*}{Spinal   Cord}       & CBCT-only   vs. proposed & 0.92            & 0.03            & 0.12            & 0.02            \\ \cline{2-6} 
			& sMRI-only   vs. proposed & 0.32            & 0.03            & 0.07            & 0.02            \\ \hline
		\end{tabular}
	\end{table}

	\subsection{Comparisons with the other methods}
	To further demonstrate the performance of our proposed method, we conducted all the experiments on the proposed method and other methods for comparison. First is the synthetic MRI deep attention U-net (sMRI DAUnet) method which was proposed in our previous study for male pelvic multi-organ (prostate, bladder, rectum) contouring\cite{RN7}. A full-image deep neural network called BibNet was proposed for CBCT male pelvic multi-organ segmentation\cite{RN1812}. And a fully convolutional deep neural network (FCDNN) was developed for CBCT mandibular canal segmentation\cite{RN1813}. Table 3 shows the DSC comparison of the results achieved by all those methods. Our proposed method outperformed other methods in all the eighteen organs for DSC comparison. In particular, compared to BibNet, our proposed method had significant improvements (p < 0.05) in DSC for organs including brain stem, left/right cochlea, left/right eye, optic chiasm, left/right optic nerve, oral cavity, left/right parotid, pharynx, and spinal cord. Except the organs including larynx and mandible, our proposed method significantly outperformed the FCDNN method. Compared to the sMRI DAUnet method, the proposed achieved significant improvements in organs including left/right eye, left/right optic nerve, oral cavity, and left/right parotid.

\begin{table}[htbp]
	\centering
	\caption{Dice similarity coefficient (DSC) comparison of the proposed, BibNet, FCDNN, and sMRI DAUnet methods.}
	\begin{tabular}{llllllll}
		\hline
		Organ             & I. Proposed & II. BibNet & III. FCDNN & \begin{tabular}[c]{@{}l@{}}IV. sMRI \\ DAUnet\end{tabular} & \begin{tabular}[c]{@{}l@{}}p value\\    \\ (I vs. II)\end{tabular} & \begin{tabular}[c]{@{}l@{}}p value\\    \\ (I vs. III)\end{tabular} & \begin{tabular}[c]{@{}l@{}}p value\\    \\ (I vs. IV)\end{tabular} \\ \hline
		Brain Stem        & 0.87±0.03   & 0.80±0.10 & 0.78±0.10  & 0.86±0.10       & \textless{}0.01                                                   & \textless{}0.01                                                   & 0.74                                                               \\ \hline
		Left Cochlea      & 0.79±0.10   & 0.75±0.09 & 0.70±0.10  & 0.76±0.09       & 0.04                                                              & \textless{}0.01                                                   & 0.15                                                               \\ \hline
		Right Cochlea     & 0.79±0.11   & 0.77±0.09 & 0.70±0.11  & 0.77±0.09       & 0.04                                                              & \textless{}0.01                                                   & 0.52                                                               \\ \hline
		Left Eye          & 0.89±0.08   & 0.84±0.15 & 0.81±0.15  & 0.84±0.15       & 0.03                                                              & \textless{}0.01                                                   & 0.02                                                               \\ \hline
		Right Eye         & 0.89±0.07   & 0.86±0.12 & 0.82±0.13  & 0.86±0.11       & 0.04                                                              & \textless{}0.01                                                   & 0.04                                                               \\ \hline
		Larynx            & 0.90±0.08   & 0.84±0.18 & 0.82±0.18  & 0.89±0.19       & 0.17                                                              & 0.09                                                              & 0.83                                                               \\ \hline
		Left Lens         & 0.75±0.06   & 0.72±0.19 & 0.61±0.15  & 0.73±0.05       & 0.09                                                              & 0.02                                                              & 0.66                                                               \\ \hline
		Right Lens        & 0.77±0.06   & 0.73±0.11 & 0.62±0.13  & 0.73±0.11       & 0.35                                                              & 0.01                                                              & 0.32                                                               \\ \hline
		Mandible          & 0.86±0.13   & 0.81±0.10 & 0.79±0.11  & 0.85±0.08       & 0.36                                                              & 0.22                                                              & 0.80                                                               \\ \hline
		Optic Chiasm      & 0.66±0.14   & 0.51±0.16 & 0.46±0.17  & 0.65±0.22       & 0.01                                                              & \textless{}0.01                                                   & 0.92                                                               \\ \hline
		Left Optic Nerve  & 0.78±0.05   & 0.69±0.09 & 0.65±0.09  & 0.71±0.07       & \textless{}0.01                                                   & \textless{}0.01                                                   & \textless{}0.01                                                    \\ \hline
		Right Optic Nerve & 0.77±0.04   & 0.69±0.09 & 0.64±0.10  & 0.69±0.09       & \textless{}0.01                                                   & \textless{}0.01                                                   & \textless{}0.01                                                    \\ \hline
		Oral Cavity       & 0.96±0.04   & 0.89±0.05 & 0.88±0.06  & 0.94±0.04       & \textless{}0.01                                                   & \textless{}0.01                                                   & \textless{}0.01                                                    \\ \hline
		Left Parotid      & 0.89±0.04   & 0.86±0.03 & 0.83±0.04  & 0.86±0.03       & \textless{}0.01                                                   & \textless{}0.01                                                   & \textless{}0.01                                                    \\ \hline
		Right Parotid     & 0.89±0.04   & 0.84±0.08 & 0.82±0.09  & 0.85±0.05       & 0.01                                                              & \textless{}0.01                                                   & \textless{}0.01                                                    \\ \hline
		Pharynx           & 0.83±0.02   & 0.64±0.15 & 0.60±0.15  & 0.78±0.14       & \textless{}0.01                                                   & \textless{}0.01                                                   & 0.28                                                               \\ \hline
		Spinal Cord       & 0.84±0.07   & 0.78±0.07 & 0.76±0.07  & 0.83±0.07       & \textless{}0.01                                                   & \textless{}0.01                                                   & 0.21                                                               \\ \hline
	\end{tabular}
\end{table}

	\bigbreak
	
	\noindent 
	\section{DISCUSSION AND CONCLUSIONS}
	
	Taking into account concurrent setup and patient anatomy, online plan adaptation has demonstrated as an encouraging approach for reducing normal tissue toxicity without delaying treatment beam delivery\cite{RN1760, RN1758}. The delineation of OARs is a necessary and important step in treatment re-planning. Manual OAR contouring is a tedious, labor-intensive, and time-consuming procedure, which is difficult to fit into current clinical workflow for re-planning process in online adaptive radiation therapy. This work aims to provide a fully automated method for directly OAR contouring on CBCT images, which are commonly acquired for patient positioning in modern radiotherapy including IMRT, VMAT, and IMPT. Extensive experiments have been carried out to demonstrate the performance of our proposed method, where, CBCT images are firstly used to generate synthetic MR images by a pre-trained cycleGAN, and the synthetic MRI and CBCT images are then fed into the dual pyramid networks for obtaining final OAR contours. Qualitatively (Figures 3-4) and quantitatively (Tables 1-2) comparisons have been conducted between our proposed method and conventional method using deep attention U-net (CBCT-only). Our proposed method outperforms the CBCT-only method in OAR contouring accuracy for majority of the organs in HN cancer patients. As shown in Tables 1 and 2, compared to the CBCT-only method, our proposed method achieved statistically significant better results in all the metrics (p < 0.05) in all the listed organs with exceptions of lenses, eyes, and mandible. For mandible, which is a bony structure, as we expected, the features for segmentation were primarily extracted from CBCT images. For lenses and eyes, our proposed method achieved comparable results with the CBCT-only method, one potential reason is that eyes or lenses in general have regular shapes and locations with respect to bony structures. 
	
	One limitation of our study is that the errors from deformable image registration might affect the performance of our proposed method. Deformable image registration was applied in two aspects in our study. Firstly, the ground truth contours on CBCT images were obtained by registration from the original contours on MRI images followed by adjustments by physicians manually. Due to the limited spatial resolution and contrast of CBCT images, it is difficult to absolutely eliminate the registration errors and obtain ideal ground truth contours that were used for training and testing the proposed method. Secondly, the cycleGAN that is used for generating synthetic MRI images was pre-trained using paired CBCT and MRI images which were spatially co-registered. The registration errors could reduce the accuracy of the model resulting in degraded image quality of synthetic MRI which to some extent affects the segmentation accuracy in our proposed method.
	
	In summary, in this study, a synthetic MRI-aided multi-organ delineation method has been implemented and validated. Through combining the features extracted from CBCT and sMRI images, our experimental results show the improvements in accuracy of our proposed method over conventional approach without sMRI aiding. To our knowledge, for the first time, a fully automated method on CBCT for HN multi-organ delineation including brain stem, cochlea, eye, larynx, lens, mandible, optic chiasm, optic nerve, oral cavity, parotid, pharynx, and spinal cord has been developed. The proposed method offers a strategic solution for OAR delineation that can be integrated into daily clinical workflow for speeding up re-planning process in plan adaptation for modern radiotherapy including IMRT, VMAT, and IMPT.

	\noindent 
	\bigbreak
	{\bf ACKNOWLEDGEMENT}
	
	This research is supported in part by the National Cancer Institute of the National Institutes of Health under Award Number R01CA215718 (XY), the Department of Defense (DoD) Prostate Cancer Research Program (PCRP) Award W81XWH-17-1-0438 (TL) and W81XWH-19-1-0567 (XD), and Emory Winship Cancer Institute pilot grant (XY).

	\noindent 
	\bigbreak
	{\bf Disclosures}
	
	The authors declare no conflicts of interest.

	\noindent 
	
	\bibliographystyle{plainnat}  
	\bibliography{arxiv}      
	
\end{document}